\documentclass[12pt]{article} 

\usepackage{tipa}
\usepackage{amsmath}
\usepackage{amssymb}
\usepackage{authblk}

\usepackage{natbib}
\usepackage{graphicx}

\newcommand{\HS}{\operatorname{HS}}
\newcommand{\HP}{\operatorname{HP}}

\newtheorem{Definition}{Definition}

\pdfoutput=1

\begin{document}

\title{A Weighted Likelihood Approach Based on Statistical Data Depths}
\author[1]{Claudio Agostinelli}
\affil[1]{Department of Mathematics, University of Trento, Trento, Italy \texttt{claudio.agostinelli@unitn.it} }
\date{\today}

\maketitle

\begin{abstract}
We propose a general approach to construct weighted likelihood estimating equations with the aim of obtain robust estimates. The weight, attached to each score contribution, is evaluated by comparing the statistical data depth at the model with that of the sample in a given point. Observations are considered regular when the ratio of these two depths is close to one, whereas, when the ratio is large the corresponding score contribution may be downweigthed. Details and examples are provided for the robust estimation of the parameters in the multivariate normal model. Because of the form of the weights, we expect that, there will be no downweighting under the true model leading to highly efficient estimators. Robustness is illustrated using two real data sets. 

\noindent \textbf{keyword}: Asymptotic Efficiency; Estimating Equations; Robustness; Statistical Data Depth; Weighted Likelihood
\end{abstract}

\section{Introduction}
Weighted Likelihood Estimating Equations (WLEE) are often used with the aim of obtaining robust estimators. \citet{green1984} is perhaps one of the earliest example, \citet{field1994} proposes a WLEE with weights that depends on the tail behavior of the distribution function, \citet{markatou1997,markatou1998} defines WLEE with weights derived from the estimating equations of a disparity minimization problems. \citet{kuchibhotla2017} further improve this approach by providing a strict connection between WLEE and disparity minimization problem. \citet{biswas2015} get ideas from both \citet{field1994} and \citet{markatou1998} and provide a similar approach based on distribution functions. Their approach is very natural and easy to implement, however one of the drawback it that the resulting estimators are not affine equivariant.

Statistical data depth was first introduced for multivariate observations, see \citet{liu2006} and the reference therein for a review. The main goal is to provide a center-outward ordering of multivariate observations that can be used for several purposes, e.g. find centers: depth maximumers; regions contains most of the observations: depth regions and quantile depth based regions; comparing distributions: Depth-Depth plot and tests based on depth, and many others. One of the main features of a depth is its invariance under affine trasformations. Furthermore, results shows that for a given model both the empirical distribution and the distribution function are completely characterized by the statistical data depth.

Our main goal is to propose a simple WLEE whose weights are based on statistical data depth. Section \ref{sec:wlee} provides an overview of the weighted likelihood stategies and our new proposal. Section \ref{sec:application} shows how we can use our new method in the multivariate normal model and Section \ref{sec:examples} illustrates the methodology with two real examples. Section \ref{sec:conclusions} reports some comments and conclusions.

\section{Weighted Likelihood based on Data Depth}
\label{sec:wlee}

Let $\mathbf{x} = (x_1, \cdots, x_n)$ be a random sample from a $p$-random vector $X$ with unknown distribution function $F$ and corresponding density function $f$. We assume a model for $X$ by $\mathcal{M} = \{M(x;\theta); \theta \in \Theta \subset \mathbb{R}^p, p \geq 1 \}$ and we denote $m_\theta(x) = m(x; \theta)$ the probability density function. Let $\hat{F}_n$ be the empirical distribution function. \citep{lindsay1994} introduced the concept of Pearson residuals $\delta$ defined by comparing the true density to the model density as 
\begin{equation*}
\delta(x) = \delta(x; M_\theta, F) = \frac{f(x)}{m(x; \theta)} - 1 \ ,
\end{equation*}
so that, when $f = m_\theta$ for a given $\theta \in \Theta$ the Pearson residuals are equal to zero for all $x$, whereas if $f \neq m_\theta$, in regions where the density $f$ is higher than $m_\theta$ the Pearson residual are large indicating the disagreement between the number of observations in that regions and the expected observations under the model. The finite sample version of the Pearson residuals is given by $\delta_n(x) = \delta(x; M_\theta, \hat{F}_n) = \frac{\hat{f}_n(x)}{m(x; \theta)} - 1$, which compares $\hat{f}_n(x)$, a non parametric estimate of $f(x)$, to the model density $m(x; \theta)$.
\citep{lindsay1994} studied a class of estimators based on the Pearson residuals for discrete models, while \citet{basu1994} and \citet{markatou1998} discussed proposals for continuous models. In particular, \citet{markatou1997,markatou1998} introduced weights evaluated by
\begin{equation}
\label{equ:weights:markatou}
w(x) = w(\delta_n(x)) = \frac{A(\delta_n(x)) + 1}{\delta_n(x) + 1} \ ,
\end{equation}
where $A(\cdot)$ is the Residual Adjustment Function \citep[RAF,][]{lindsay1994,park2002} obtaining a Weighted Likelihood Estimating Equations (WLEE)
\begin{equation}
\label{equ:wlee}
\frac{1}{n} \sum_{i=1}^n w(x_i;M_\theta,\hat{F}_n) \ u(x_i; \theta) \ ,
\end{equation}
where $u(y_i; \theta)$ denotes the $i$-th contribution to the score function. These estimating equations are derived from a disparity measure however, there is no exact link between the two approaches. Recently, \citet{kuchibhotla2017} provide a WLEE in the same spirit which corresponds to a disparity measures, they also formally prove the asympotic and robust properties of their approach. In the attempt to avoid the use of non parametric estimators \citet{biswas2015} proposes to define Pearson residuals as ratio of distribution functions such as
\begin{equation*}
\delta_n(x) = 
\begin{cases}
\frac{\hat{F}_n(x)}{M(x;\theta)} - 1\ , & \text{if} \ M(x;\theta) \le 0.5 \\
\frac{1 - \hat{F}_n(x)}{1 - M(x;\theta)} - 1\ , &  \text{if} \ M(x;\theta) > 0.5 \\
\end{cases}
\end{equation*}
Let $H(\delta)$ be a smooth function defined on $[-1,\infty)$, which assumes its maximum value $1$ at $\delta=0$ and descends smoothly in either tail as $\delta$ moves away from $0$, i.e., $H(0) = 1$ and $H'(0) = 0$ and the next higher non-zero derivative at $\delta=0$ has a negative sign; an example is 
\begin{equation*}
H(\delta, a, c) = 
\begin{cases}
\exp(-a \delta^2) \ , & \text{if} \ \delta \le c \\
0 & \text{otherwise}
\end{cases} 
\end{equation*}
where $a > 0$ and $c > 0$ are constants. \citet{biswas2015} defines a weight function as
\begin{equation}
\label{equ:weights}
w(x) = 
\begin{cases}
1 \ , & \text{if} \ p < M(x,\theta) < 1 - p \\
H(\delta, a, \infty) \ , & \text{otherwise.} \nonumber
\end{cases}
\end{equation}
Clearly, also weights in the form of (\ref{equ:weights:markatou}) are possible. Their approach is general and can be used in multivariate setting as they show a bivariate example. However, their definition makes the Pearson residuals not affine invariant and hence the final estimates are not affine equivariant. We are going to propose a general approach to construct weights in the same spirit using statistical data depth which is more natural and is affine invariant.

Let $D(x; F_X)$ be a statistical data depth \citep{zuo2000a,liu2006} for the point $x \in \mathbb{R}^p$ according to the distribution $F_X$ of the r.v. $X \in \mathbb{R}^p$. Let $D_n(x; \hat{F}_n)$ be the finite sample version based on the empirical distribution function $\hat{F}_n$ of the sample $\mathbf{x}$. Denote $\mathcal{F}$ the class of distributions in $\mathbb{R}^p$, assume that $D(x; F_X)$ satisfies the following properties \citep{liu1990,zuo2000a}
\begin{description}
\item[P1.] \textit{Affine Invariance}. $D(A x + b; F_{A X + b}) = D(x; F_X)$ for any distribution function $F_X \in \mathcal{F}$, any $p \times p$ nonsingular matrix $A$ and any $p$-vector $b$. 
\item[P2.] \textit{Maximality at Center}. For an $F_X$ having ``center'' $\mu$ (e.g. the point of symmetry relative to some notion of symmetry), $D(\mu;F_X) = \sup_{x \in \mathbb{R}^p} D(x; F_X)$.
\item[P3.] \textit{Monotonicity Relative to Deepest Point}. For any $F_X$ having deepest point $\mu$ (i.e., point of maximal depth), $D(x; F_X) \le D(\mu + \alpha (x - \mu); F_X)$, $\alpha \in [0,1]$.
\item[P4.] \textit{Vanishing at Infinity}. $D(x; F_X) \rightarrow 0$ as $||x|| \rightarrow \infty$, for each $F_X \in \mathcal{F}$.
\end{description}
We define the Pearson residual as
\begin{equation*}
\delta(x) = \frac{D(x; F)}{D(x;M_\theta)} - 1
\end{equation*}
and the finite sample version as
\begin{equation*}
\delta_n(x) = \frac{D_n(x; \hat{F}_n)}{D(x;M_\theta)} - 1
\end{equation*}
This Pearson residual have the desired behaviour being equal to $0$ whenever $F = M_\theta$ for some $\theta \in \Theta$, and of attaining large values in regions where the two distributions are mismatched; furthermore, because of the invariance property \textbf{P1.} of $D$, the Pearson residual is also invariant to affine transformations. For most depths a uniform convergence of $D_n(x; \hat{F}_n)$ to $D(x; F)$ holds almost surely; hence we expect that, at the model, the proposed method is highly efficient. Using weights based on (\ref{equ:weights:markatou}) or (\ref{equ:weights}) leads to a WLEE that can be solved by an iterative reweigthing algorithm.

\section{Application to the Multivariate Normal model}
\label{sec:application}
Consider $M_\theta$ be a Multivariate Normal model where $\theta = (\mu, \Sigma)$, $\mu$ is the mean vector and $\Sigma$ is the variance-covariance matrix. We discuss how to evaluate the proposed Pearson residual using the halfspace depth. We first review the concept of halfspace depth \citep{tukey1975,donoho1992}. Let $\HS_{u}(a)$ be the closed halfspace $\left\{z \in \mathbb{R}^p: u^\top z \geq a \right\}$. Here $u$ is a $p$-vector satisfying $u^\top u = 1$. Note that $\HS_{u}(a)$ is the positive side of the hyperplane $\HP_{u}(a) = \left\{ z \in \mathbb{R}^p : u^\top z = a \right\}$. The negative side $\HS_{-u}(a)$ of $\HP_{u}(a)$ is similarly defined. Halfspace depth is the minimum probability of all halfspaces including $x$.
\begin{Definition}
The halfspace depth $d_{HS}(x;X)$ maps $x \in \mathbb{R}^p$ to the minimum probability, according to the random vector $X$, of all closed halfspaces including $x$, that is
\begin{align*}
d_{HS}(x; X) & = \inf_{u: u^\top u=1} \Pr(\HS_{u}(u^\top x); X) \\
& = \inf_{u: u^\top u=1} \Pr( z \in \mathbb{R}^p : u^\top z \geq u^\top x; X) \ .
\end{align*}
\end{Definition}
This statistical data depth is particularly usefull since its properties. In particular, \citet[][Theorem 1]{struyf1999} show that the finite sample halfspace depth characterizes the empirical distribution. \citet[][Corollary 3.1]{kong2010} show that the halfspace depth characterizes the underlying distribution. Furthermore, \citet[][Theorem 3.3, Corollay 4.3]{zuo2000b} shows that the halfspace depth for a multivariate normal model can be easily obtained since $D(x; M_\theta) = (1 - F_{\chi^2_p}(d(x;\theta)))/2$ where $d(x, \theta) = (x - \mu)^\top \Sigma^{-1}(x - \mu)$ is the squared Mahalanobis distance and $F_{\chi^2_p}$ is the distribution function of a $\chi^2_p$ chi-squared with $p$ degrees of freedom random variable. Finally, calculation of the finite sample halfspace depth in a dimension $p$ can be performed efficiently using the algorithms proposed in \citet{liu2017} and \citet{dyckerhoff2016} and available in the \texttt{R} package \texttt{ddalpha} by \citet{pokotylo2016}. Similar results are available for any model belonging to the elliptically symmetric family of distributions.

\section{Examples}
\label{sec:examples}
\begin{figure}
\centering
\includegraphics[width=0.45\textwidth]{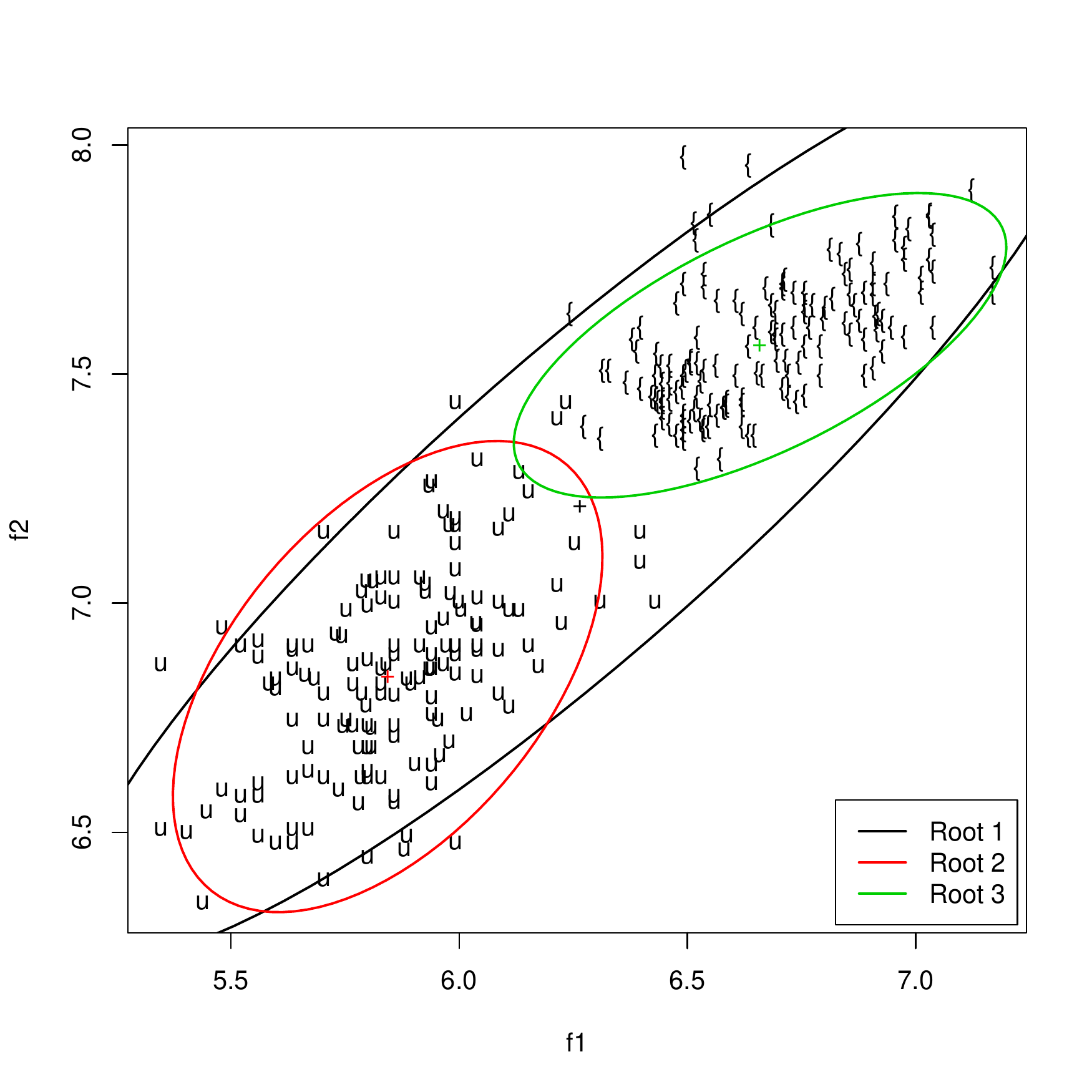}
\includegraphics[width=0.45\textwidth]{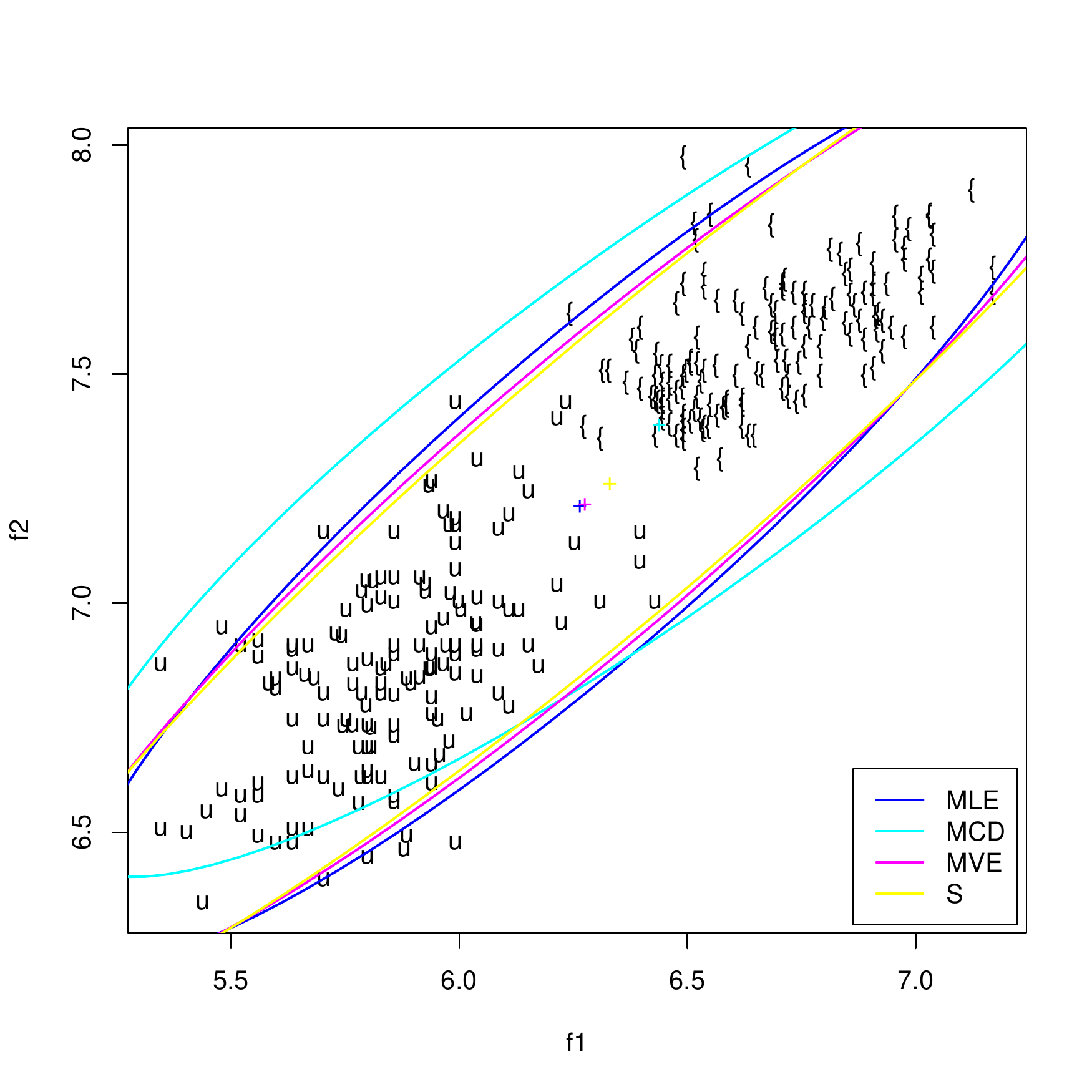}
\caption{Vowel Recognition Data. Bivariate example, vowels ``u'' and ``\ae''. Left: the three roots of the proposed method. Right: estimates provived by MLE and robust procedures. Ellipses are $95\%$ regions.}
\label{fig:x1}
\end{figure}

We consider the data set \texttt{pb52} available in the \texttt{R} package \texttt{phonTools} \citet{barreda2015} which contains the Vowel Recognition Data considered in \citet{peterson1952}, see also \citet{boersma2012}. In a first example a bivariate data set is illustrated, where we consider the vowels ``u'' (close back rounded vowel) and ``\ae'' (near-open front unrounded vowel, ``\{'' as x-sampa symbol) with a sample of size $304$ equally divided for each vowel and the log transformed F1 and F2 frequencies measured in Hz. Our procedure use the following settings $\alpha=0.5$, $a=0.05$, $c=200$, and uses $500$ subsamples of size $6$ as starting values for finding the roots. We also consider the Maximum Likelihood (MLE), the Minimum Covariance Determinant (MCD), the Minimum Volume Ellipsoid (MVE) and the S-Estimates (S) as implemented in the \texttt{R} package \texttt{rrcov} by \citet{todorov2009}, the last three procedures were used with the exaustive subsampling explorations.  Figure \ref{fig:x1} in the left panel reports the three estimates found by our methods. While the first root coincides with the MLE, the other two nicely identify the two subgroups. This is not the case for all the other investigated methods, see Figure \ref{fig:x1} right panel, where the estimates are approximately all coincident with the MLE.

\begin{figure}
\centering
\includegraphics[width=0.45\textwidth]{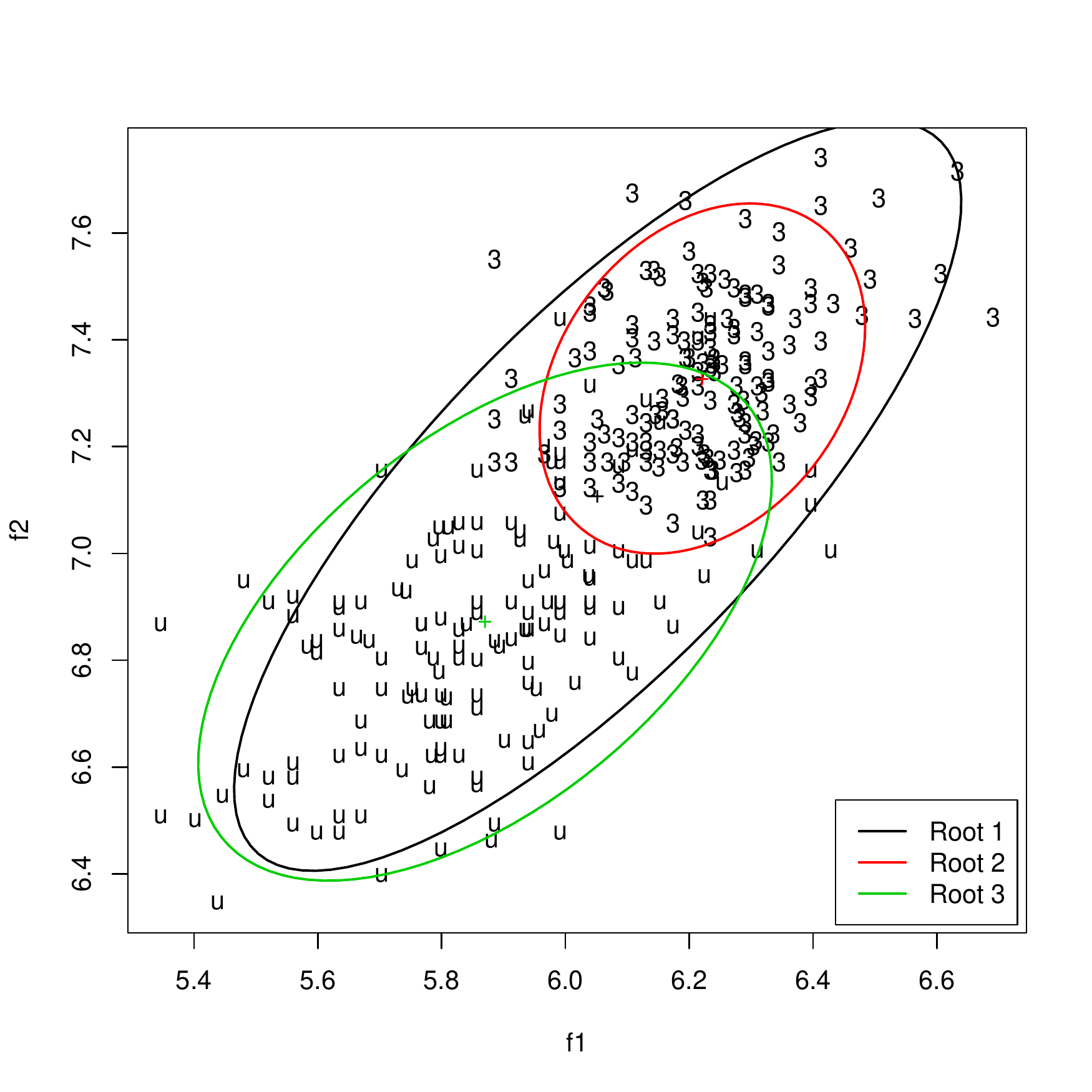}
\includegraphics[width=0.45\textwidth]{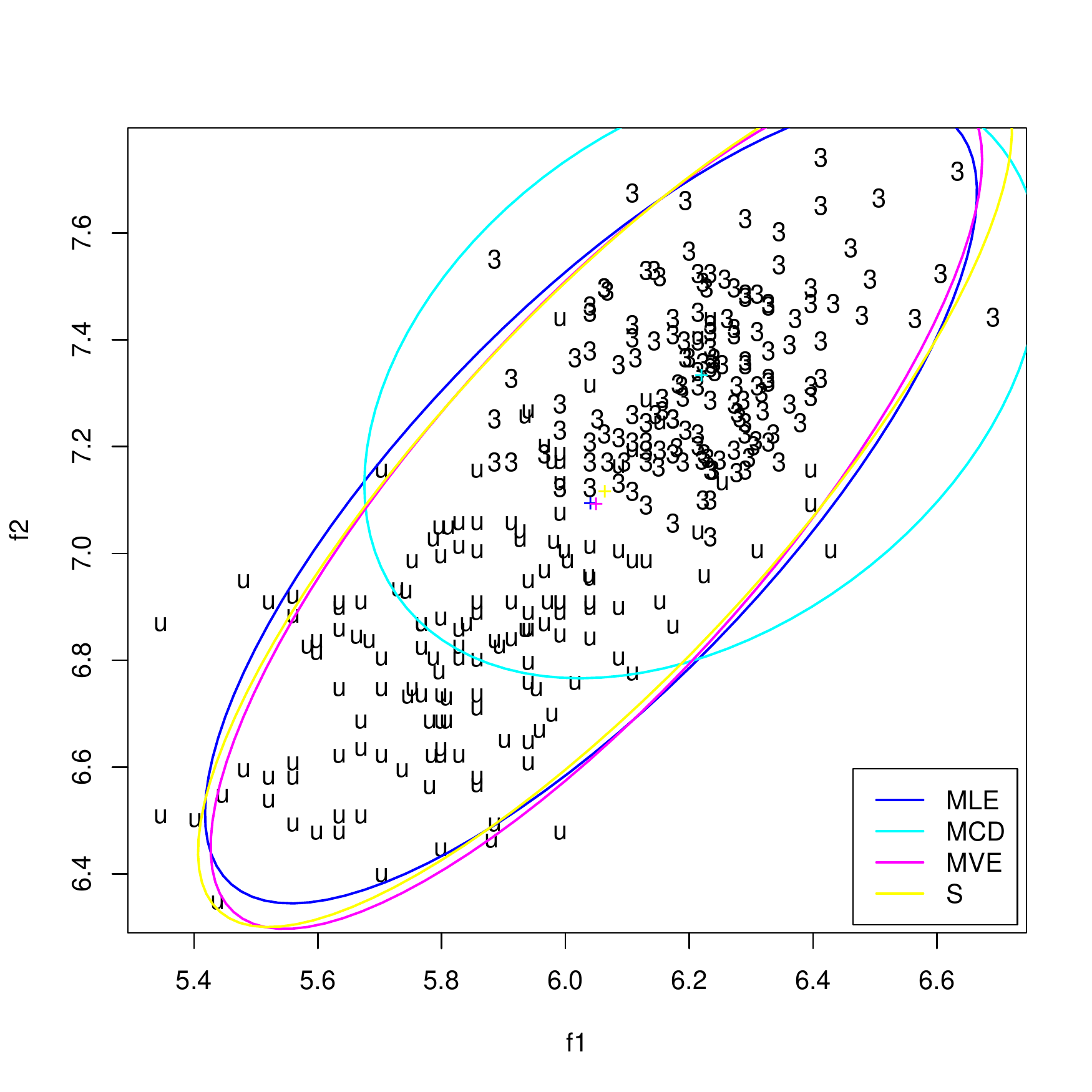} \\
\includegraphics[width=0.45\textwidth]{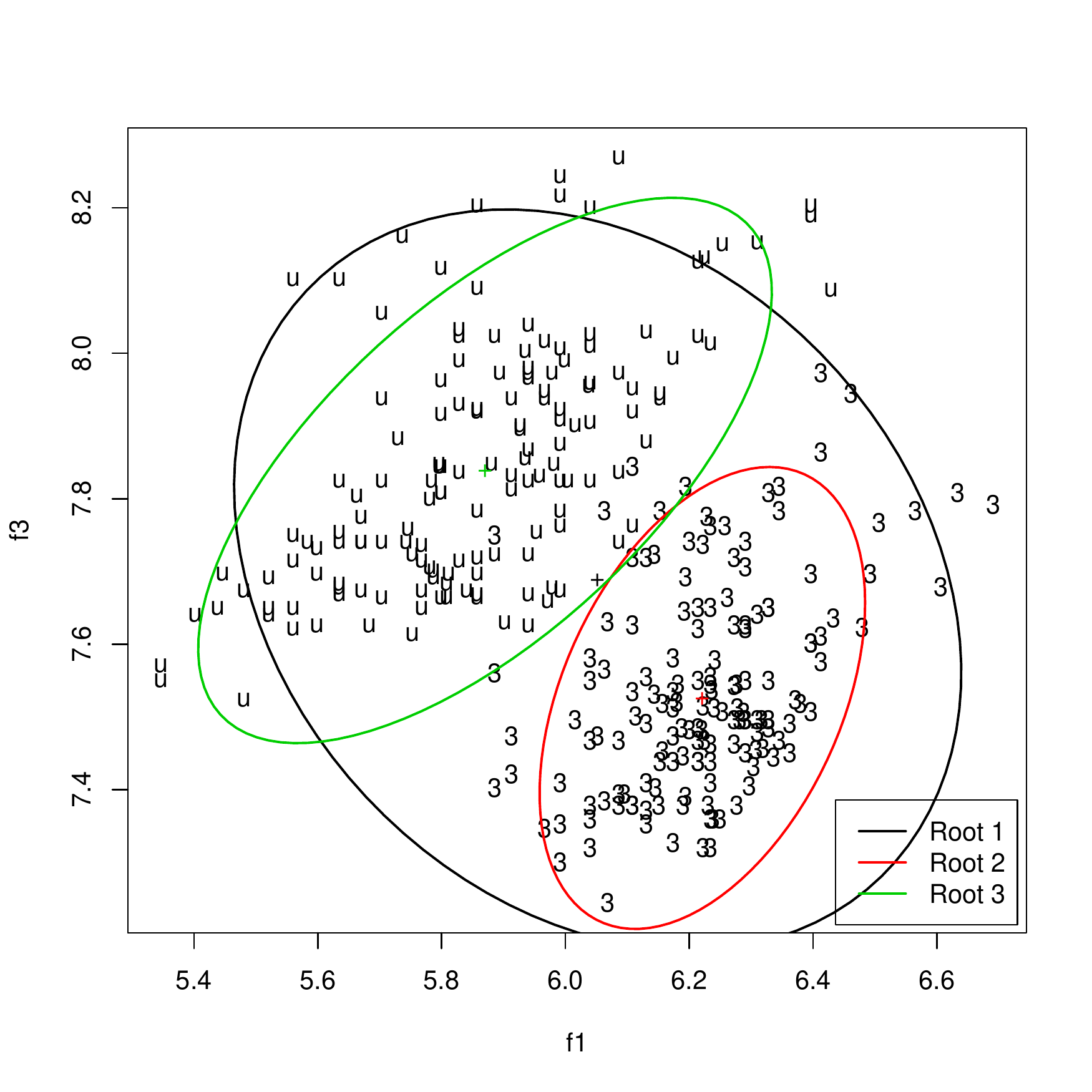}
\includegraphics[width=0.45\textwidth]{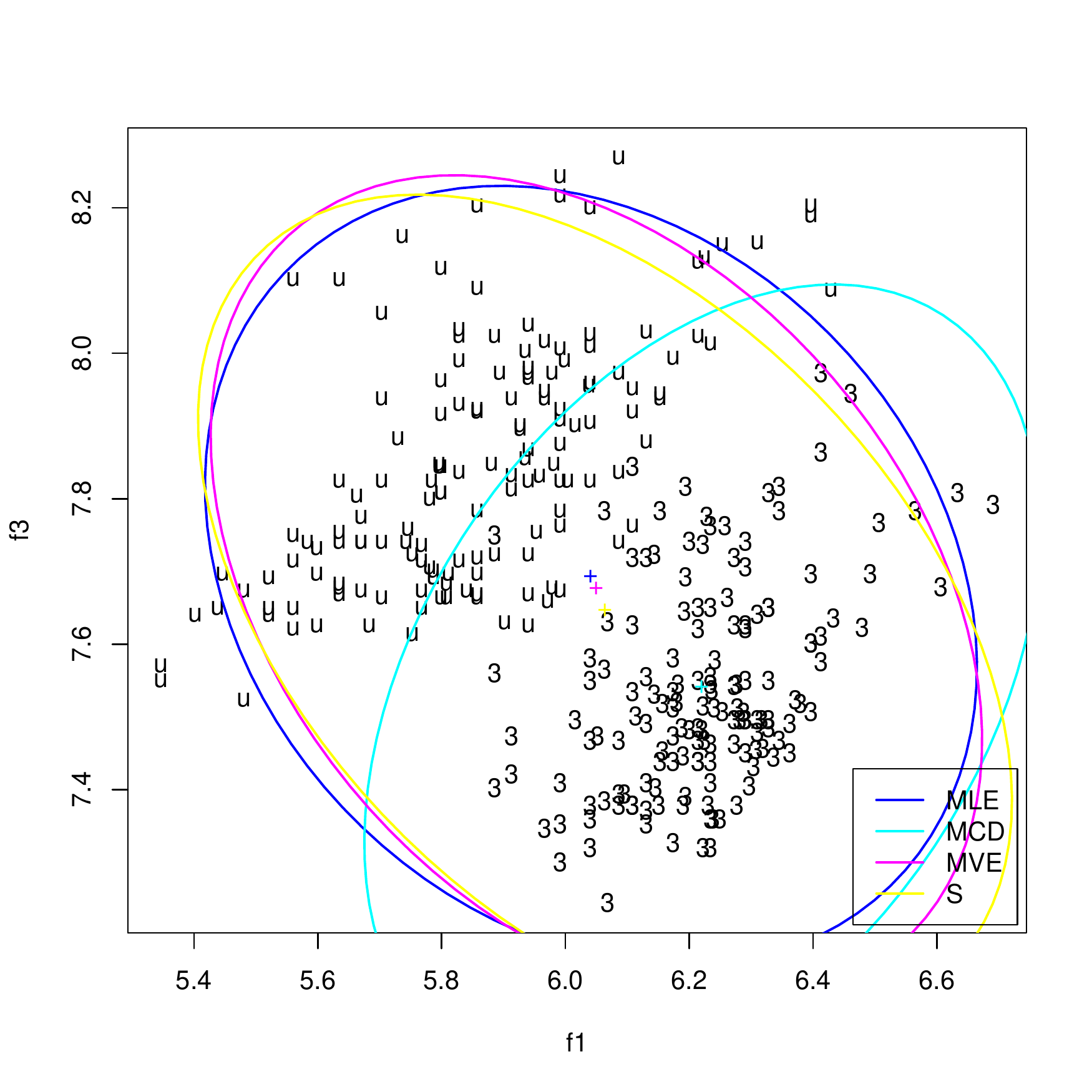} \\
\includegraphics[width=0.45\textwidth]{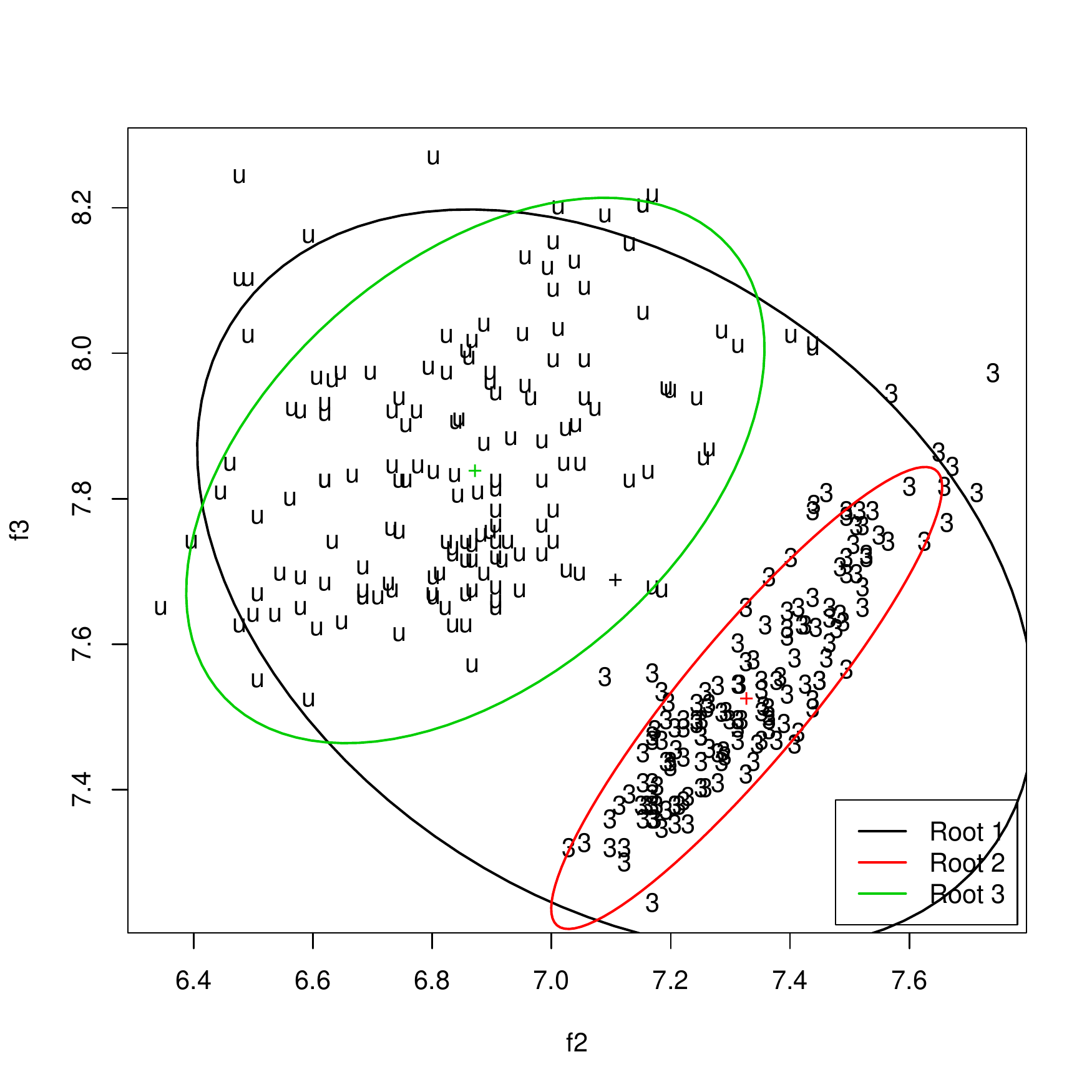}
\includegraphics[width=0.45\textwidth]{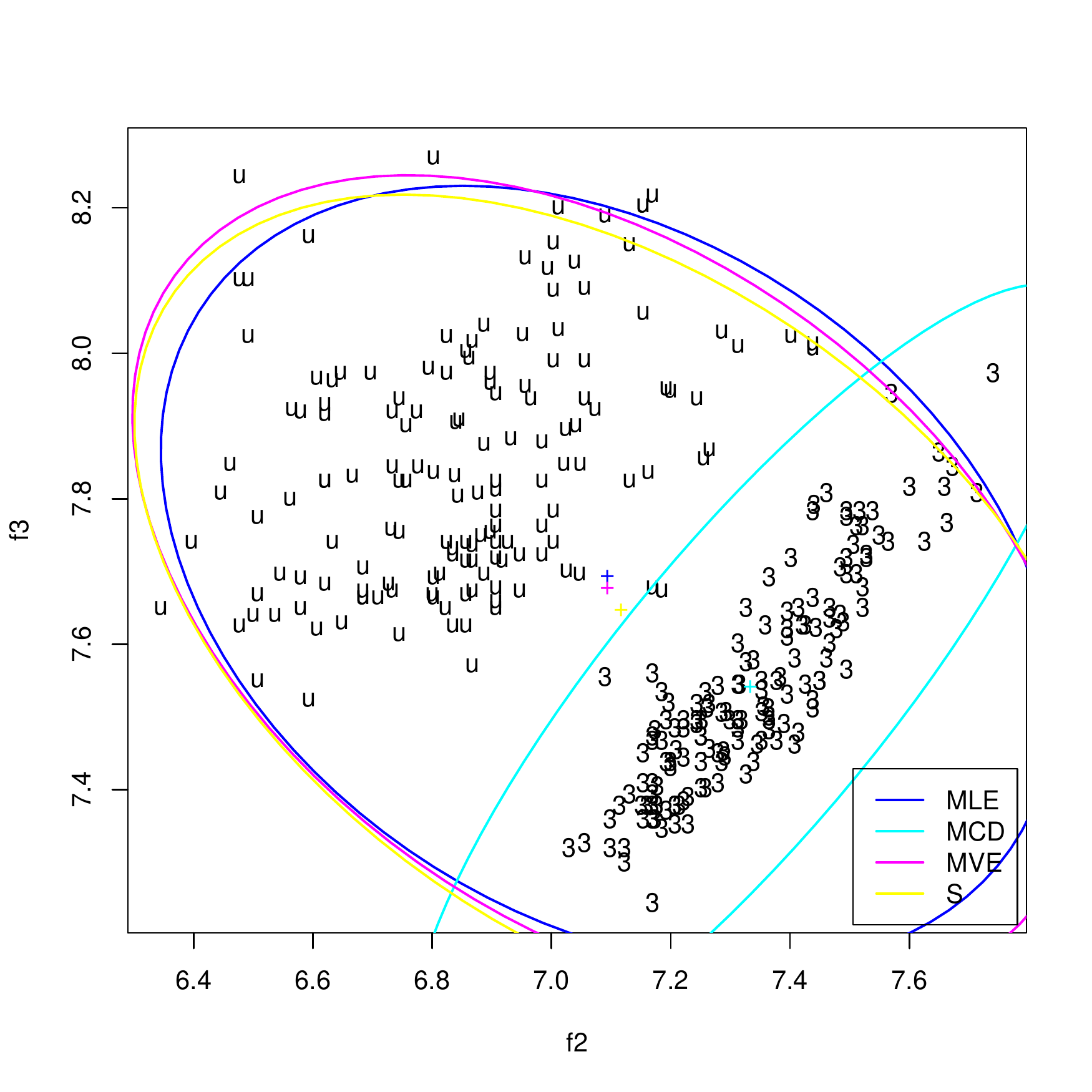} \\
\caption{Vowel Recognition Data. Trivariate example, vowels ``u'' and ``\textrhookrevepsilon''. Left: the three roots of the proposed method. Right: estimates provived by MLE and robust procedures. Ellipses are $95\%$ regions.}
\label{fig:x3}
\end{figure}

In a second example a trivariate data set is used considering vowels ``u'' (close back rounded vowel) and ``\textrhookrevepsilon'' (open-mid central unrounded vowel, ``3'' or ``$3^\backprime$'' as x-sampa symbol) with a sample of size $304$ equally divided for each vowel and the log transformed F1, F2 and F3 frequencies measured in Hz. Figure (\ref{fig:x3}) shows the results. For the classical robust procedures only MCD is able to somehow recover the structure of the observations for the vowels ``\textrhookrevepsilon'', while the other behave like the MLE. Our procedure finds nicely all the two substructure. For this data our procedure was set using  $\alpha=0.25$, $a=0.1$, $c=30$, and $1000$ subsamples of size $6$ as starting values for finding the roots. Using the setting of the first examples only the first two roots are found.

\section{Conclusions}
\label{sec:conclusions}
We have outlined a new form of weighted likelihood estimating equations where weights are based on comparing statistical data depth of the sample with that of the model. This approach avoids the use of nonparametric density estimates which can lead to problems for multivariate data, while retains nice characteristics of the classical WLEE approach, that are high efficiency at the model, affine equivariance and robustness. In the future, we hope to formally establish the theoretical properties of the proposed estimator.


\begin{thebibliography}{24}
\providecommand{\natexlab}[1]{#1}
\providecommand{\url}[1]{\texttt{#1}}
\expandafter\ifx\csname urlstyle\endcsname\relax
  \providecommand{\doi}[1]{doi: #1}\else
  \providecommand{\doi}{doi: \begingroup \urlstyle{rm}\Url}\fi

\bibitem[Barreda(2015)]{barreda2015}
S.~Barreda.
\newblock \emph{phonTools: Functions for phonetics in R.}, 2015.
\newblock R package version 0.2-2.1.

\bibitem[Basu and Lindsay(1994)]{basu1994}
A.~Basu and B.G. Lindsay.
\newblock Minimum disparity estimation for continuous models: efficiency,
  distributions and robustness.
\newblock \emph{Annals of the Institute of Statistical Mathematics},
  46\penalty0 (4):\penalty0 683--705, 1994.

\bibitem[Biswas et~al.(2015)Biswas, Roy, Majumder, and Basu]{biswas2015}
A.~Biswas, T.~Roy, S.~Majumder, and A.~Basu.
\newblock A new weighted likelihood approach.
\newblock \emph{Stat}, 4\penalty0 (1):\penalty0 97--107, 2015.

\bibitem[Boersma and Weenink(2012)]{boersma2012}
P.~Boersma and D.~Weenink.
\newblock Praat: doing phonetics by computer [computer program]. version
  5.3.19.
\newblock http://www.praat.org/, 2012.
\newblock retrieved 24 June 2012.

\bibitem[Donoho and Gasko(1992)]{donoho1992}
D.L. Donoho and M.~Gasko.
\newblock Breakdown properties of location estimates based on halfspace depth
  and projected outlyingness.
\newblock \emph{The Annals of Statistics}, 20:\penalty0 1808--1827, 1992.

\bibitem[Dyckerhoff and Mozharovskyi(2016)]{dyckerhoff2016}
R.~Dyckerhoff and P.~Mozharovskyi.
\newblock Exact computation of the halfspace depth.
\newblock \emph{Computational Statistics \& Data Analysis}, 98:\penalty0
  19--30, 2016.

\bibitem[Field and Smith(1994)]{field1994}
C.~Field and B.~Smith.
\newblock Robust estimation -- a weighted maximum likelihood approach.
\newblock \emph{International Statistical Review}, 62:\penalty0 405--424, 1994.

\bibitem[Green(1984)]{green1984}
P.J. Green.
\newblock Iteratively reweighted least squares for maximum likelihood
  estimation, and some robust and resistent alternatives.
\newblock \emph{Journal of the Royal Statistical Society: Series B},
  46:\penalty0 149--192, 1984.

\bibitem[Kong and Zuo(2010)]{kong2010}
L.~Kong and Y.~Zuo.
\newblock Smooth depth contours characterize the underlying distribution.
\newblock \emph{Journal of Multivariate Analysis}, 101:\penalty0 2222--2226,
  2010.

\bibitem[Kuchibhotla and Basu(2017)]{kuchibhotla2017}
A.K. Kuchibhotla and A.~Basu.
\newblock A minimum distance weighted likelihood method of estimation.
\newblock Technical report, Interdisciplinary Statistical Research Unit (ISRU),
  Indian Statistical Institute, Kolkata, India, 2017.

\bibitem[Lindsay(1994)]{lindsay1994}
B.G. Lindsay.
\newblock Efficiency versus robustness: The case for minimum hellinger distance
  and related methods.
\newblock \emph{The Annals of Statistics}, 22:\penalty0 1018--1114, 1994.

\bibitem[Liu(1990)]{liu1990}
R.Y. Liu.
\newblock On a notion of data depth based on random simplices.
\newblock \emph{The Annals of Statistics}, 18\penalty0 (1):\penalty0 405--414,
  1990.

\bibitem[Liu et~al.(2006)Liu, Serfling, and Souvaine]{liu2006}
R.Y. Liu, R.J. Serfling, and D.L. Souvaine.
\newblock \emph{Data depth: robust multivariate analysis, computational
  geometry, and applications}.
\newblock {AMS} Bookstore, 2006.

\bibitem[Liu(2017)]{liu2017}
X.~Liu.
\newblock Fast implementation of the tukey depth.
\newblock \emph{Computational Statistics}, 32\penalty0 (4):\penalty0
  1395--1410, 2017.

\bibitem[Markatou et~al.(1997)Markatou, Basu, and Lindsay]{markatou1997}
M.~Markatou, A.~Basu, and B.G. Lindsay.
\newblock Weighted likelihood estimating equations: the discrete case with
  applications to logistic regression.
\newblock \emph{Journal of Statistical Planning and Inference}, 57:\penalty0
  215--232, 1997.

\bibitem[Markatou et~al.(1998)Markatou, Basu, and Lindsay]{markatou1998}
M.~Markatou, A.~Basu, and B.G. Lindsay.
\newblock Weighted likelihood equations with bootstrap root search.
\newblock \emph{Journal of the American Statistical Association}, 93\penalty0
  (442):\penalty0 740--750, 1998.

\bibitem[Park et~al.(2002)Park, Basu, and Lindsay]{park2002}
C.~Park, A.~Basu, and B.G. Lindsay.
\newblock The residual adjustment function and weighted likelihood: a graphical
  interpretation of robustness of minimum disparity estimators.
\newblock \emph{Computational Statistics \& Data Analysis}, 39\penalty0
  (1):\penalty0 21--33, 2002.

\bibitem[Peterson and Barney(1952)]{peterson1952}
G.E. Peterson and H.L. Barney.
\newblock Control methods used in a study of the vowels.
\newblock \emph{Journal of the Acoustical Society of America}, 24:\penalty0
  175--184, 1952.

\bibitem[Pokotylo et~al.(2016)Pokotylo, Mozharovskyi, and
  Dyckerhoff]{pokotylo2016}
O.~Pokotylo, P.~Mozharovskyi, and R.~Dyckerhoff.
\newblock Depth and depth-based classification with r-package ddalpha.
\newblock \emph{arXiv:1608.04109}, 2016.

\bibitem[Struyf and Rousseeuw(1999)]{struyf1999}
A.~Struyf and P.J. Rousseeuw.
\newblock Halfspace depth and regression depth characterize the empirical
  distribution.
\newblock \emph{Journal of Multivariate Analysis}, 69:\penalty0 135--153, 1999.

\bibitem[Todorov and Filzmoser(2009)]{todorov2009}
V.~Todorov and P.~Filzmoser.
\newblock An object-oriented framework for robust multivariate analysis.
\newblock \emph{Journal of Statistical Software}, 32\penalty0 (3):\penalty0
  1--47, 2009.

\bibitem[Tukey(1975)]{tukey1975}
J.W. Tukey.
\newblock Mathematics and picturing of data.
\newblock In \emph{Proceedings of International Congress of Mathematics},
  volume~2, pages 523--531, 1975.

\bibitem[Zuo and Serfling(2000{\natexlab{a}})]{zuo2000a}
Y.~Zuo and R.J. Serfling.
\newblock General notions of statistical depth function.
\newblock \emph{The Annals of Statistics}, 28\penalty0 (2):\penalty0 461--482,
  2000{\natexlab{a}}.

\bibitem[Zuo and Serfling(2000{\natexlab{b}})]{zuo2000b}
Y.~Zuo and R.J. Serfling.
\newblock Structual properties and convergence results for contours of sample
  statistical depth functions.
\newblock \emph{The Annals of Statistics}, 28\penalty0 (2):\penalty0 483--499,
  2000{\natexlab{b}}.

\end{thebibliography}

\end{document}